\documentclass{ifacconf}

\makeatletter
\let\old@ssect\@ssect 
\makeatother

\usepackage[nolist]{acronym}
\usepackage{xcolor}
\usepackage{amsmath}
\usepackage{amssymb}
\usepackage{mathtools}
\usepackage{hyperref}
\usepackage{mathrsfs}

\newcommand{\R}{\mathbb{R}} 
\newcommand{\K}{\mathbb{K}} 
 
\newcommand{\N}{\mathbb{N}}

\newcommand{\dc}{\mathcal{D}}

\newcommand{\M}{\mathbb{M}}

\newcommand{\0}{\mathbf{0}}
\newcommand{\1}{\mathbf{1}}

\DeclarePairedDelimiter{\abs}{\lvert}{\rvert}
\DeclarePairedDelimiter{\norm}{\lVert}{\rVert}

\DeclarePairedDelimiter{\rank}{\textrm{rank}(}{)}

\DeclarePairedDelimiterX{\inp}[2]{\langle}{\rangle}{#1, #2}

\DeclarePairedDelimiter{\Mp}{\mathcal{M}_+(}{)}

\usepackage{natbib}        

\makeatletter
\def\@ssect#1#2#3#4#5#6{%
  \NR@gettitle{#6}
  \old@ssect{#1}{#2}{#3}{#4}{#5}{#6}
}
\makeatother
\usepackage{color,soul}
\usepackage{comment}
\begin{document}
\begin{frontmatter}

\title{Frequency-Domain  Identification of Discrete-Time Systems using Sum-of-Rational Optimization\thanksref{footnoteinfo}} 

\thanks[footnoteinfo]{This work was partially supported by the Swiss National Science Foundation under NCCR Automation, grant agreement 51NF40\_180545.}

\author[First]{Mohamed Abdalmoaty} 
\author[First]{Jared Miller} 
\author[First]{Mingzhou Yin} 
\author[First]{Roy Smith} 

\address[First]{ Automatic Control Laboratory (IfA) and NCCR Automation, Department of Information Technology and Electrical Engineering (D-ITET), ETH Z\"{u}rich, Physikstrasse 3, 8092, Z\"{u}rich, Switzerland \\(e-mail: \{mabdalmoaty, jarmiller, myin, rsmith\}@control.ee.ethz.ch).}

\begin{abstract}
\label{sec:abstract}
We propose a computationally tractable method for the identification of stable canonical discrete-time rational transfer function models, using frequency domain data. The problem is formulated as a global non-convex optimization problem whose objective function is the sum of weighted squared residuals at each observed frequency datapoint. Stability is enforced using a polynomial matrix inequality constraint.  The problem is solved \textit{globally} by a moment-sum-of-squares hierarchy of semidefinite programs through a framework for sum-of-rational-functions optimization. 
Convergence of the moment-sum-of-squares program is guaranteed as the bound on the degree of the sum-of-squares polynomials approaches infinity. The performance of the proposed method is demonstrated using  numerical simulation examples. 




\end{abstract}


\end{frontmatter}

\begin{abstract}
\label{sec:abstract}
We propose a computationally tractable method for the identification of stable canonical discrete-time rational transfer function models, using frequency domain data. The problem is formulated as a global non-convex optimization problem whose objective function is the sum of weighted squared residuals at each observed frequency datapoint. Stability is enforced using a polynomial matrix inequality constraint.  The problem is solved \textit{globally} by a moment-sum-of-squares hierarchy of semidefinite programs through a framework for sum-of-rational-functions optimization. 
Convergence of the moment-sum-of-squares program is guaranteed as the bound on the degree of the sum-of-squares polynomials approaches infinity. The performance of the proposed method is demonstrated using  numerical simulation examples. 




\end{abstract}
\section{Introduction}
\label{sec:introduction}

Fitting parametric \ac{LTI} models to noisy frequency domain data is a classical problem in system identification \citep{pintelon2012, ljung1999}. Many methods and algorithms based on stochastic as well as deterministic frameworks have been developed and used. In particular, the \ac{ML} method, and \acp{PEM} are frequently used due to their optimal asymptotic statistical properties; see \cite[Ch.~8 \& 9]{ljung1999} and \cite[Ch. 9.11]{pintelon2012}. They define estimators as solutions to optimization problems; however, except for a few very specific model structures (e.g., ARX models), their implementation is complicated by the fact that the resulting optimization problems are highly non-convex.  Solutions of \acp{PEM}/\ac{ML} are usually obtained by local numerical optimization algorithms \citep{Wills_2008}, introducing the vulnerability of terminating at a local minimum without a guarantee of global optimally.

The goal of this paper is to perform parametric identification of stable canonical rational \ac{LTI} models through global optimization of a weighted 2-norm of the residuals in the frequency domain. For  strictly proper discrete-time SISO system with a known fixed order $n$ parameterized by $(a, b) \in \R^n \times \R^n$,  the model is given in terms of a transfer function 
\begin{align} \label{eq:transfer_function}
    G(z; (a,b)) = \frac{B(z^{-1})}{1+A(z^{-1})} = \frac{\sum_{k=1}^{n} b_k z^{-k}}{1 + \sum_{k=1}^n a_k z^{-k}}
\end{align}
where $z$ is the complex variable of the $\mathcal{Z}$-transform. 
The problem is formulated in the frequency domain using a sequence of 
$N_f$ corrupted Bode-plot data $G_f$ at frequencies $\omega_f$, forming the data set $\dc =\{(\omega_f, G_f)\}_{f=1}^{N_f}$. The data in $\dc$ can come from an empirical transfer function estimate, or swept-sine experiments. The following observation process is considered with residual $\eta_f$:
\begin{align}
    G_f = G_\circ(e^{j \omega_f}) + \eta_f. \label{eq:gain_datas}
\end{align}
where $G_\circ(e^{j \omega_f}) = G(e^{j \omega_f}; (a_\circ, b_\circ))$ denotes the true underlying system. When $G_\circ$ is replaced with $G\left(e^{j \omega_f}; (a,b)\right)$ the following weighted 2-norm of the residuals vector $\eta(a,b)$ is obtained
\begin{equation}
\begin{aligned}
    \mathcal{J}_\dc (a, b) &= \textstyle \norm{W \odot \eta(a, b)}_2^2 \\
    &= \textstyle \sum_{f=1}^{N_f}  \abs{W_f\left(G_f - G\left(e^{j \omega_f}; (a,b)\right)\right)}^2. \label{eq:mse_plant}
\end{aligned}
\end{equation}
where $W = \{W_f\}_{f=1}^{N_f}$ are user-defined weights. Our goal is to identify a model $(a^*, b^*)$ that \textit{globally} infimizes  the weighted 2-norm of the residuals vector: 
\begin{align}
    P^* =  \inf_{(a, b)}  \mathcal{J}_\dc (a, b).\label{eq:gain_base}
\end{align}

Problem \eqref{eq:gain_base} is non-convex in the optimization variables $(a, b)$. The objective function can be cast as a sum of rational functions (sum of ratios) problem \citep{schaible2003fractional, freund2001solving}. 
Our work focuses on global optimization of \eqref{eq:gain_base} through the measure \ac{LP}-based sum-of-ratios framework introduced by \cite{bugarin2016minimizing}. The method in \cite{bugarin2016minimizing} uses the moment-\ac{SOS} \citep{lasserre2009moments} hierarchy of \acp{SDP} in order to obtain a nondecreasing sequence of lower bounds to $P^*$. Recovery of global optimizers $(a^*, b^*)$ will occur if the \ac{PSD} solution matrices satisfy a rank condition \citep{curto1996solution, henrion2005detecting}; a global optimality certificate can then be obtained.

\subsection{Contributions}
Earlier works that use the moment-\ac{SOS} hierarchy for parametric identification in time-domain are \cite{Rodrigues2019,Rodrigues2020}. These approaches are based on introducing the time-domain residuals as optimization variables, leading to sparse polynomial optimization problems. 
A related prior work to our method is in Ch. 7-9 of \cite{vuillemin2014frequency}, in which model reduction of an \ac{LTI} system proceeds through a sum-of-ratios function optimization problem by minimizing a norm over a pole-residue partial fraction expansion. 

The contributions of this research are:
\begin{itemize}
    \item A first application of the sum-of-rational framework of \cite{bugarin2016minimizing} to parametric frequency-domain system identification.
    \item The stability of the identified models is enforced using  a Polynomial Matrix Inequality (PMI) constraint.
    \item An \ac{LP} in measures that admits a convergent sequence of truncations whose objectives converge to the global minimum of the original objective function. 
\end{itemize}

The paper is organized as follows. Section \ref{sec:preliminaries} introduces the notation used, reviews the moment-\ac{SOS} hierarchy of \acp{SDP} and the sum-of-rational framework of \cite{bugarin2016minimizing}. Section \ref{sec:rational_sysid} casts the global optimization problem \eqref{eq:gain_base} as a real-valued sum-of-rational optimization problem. Section \ref{sec:mom_sdp} formulates finite-degree \ac{LMI} truncations of the sum-of-rational \acp{LP} and tabulates their \ac{SDP}-based computational complexity. 
Section \ref{sec:examples} demonstrates the performance of the proposed method using numerical simulation examples. 
Section \ref{sec:conclusion} concludes the paper. An extended version of this paper can be found in \cite{abdalmoaty2023rationalsysid}.

\section{Preliminaries}
\label{sec:preliminaries}

\begin{acronym}[PEM]

\acro{BSA}{Basic Semialgebraic}



\acro{LTI}{Linear Time-Invariant}
\acroindefinite{LTI}{an}{a}

\acro{LMI}{Linear Matrix Inequality}
\acroplural{LMI}[LMIs]{Linear Matrix Inequalities}
\acroindefinite{LMI}{an}{a}


\acro{LP}{Linear Program}
\acroindefinite{LP}{an}{a}

\acro{ML}{Maximum Likelihood}

\acro{MSE}{Mean Squared Error}


\acro{PEM}{Prediction Error Method}

\acro{PMI}{Polynomial Matrix Inequality}


\acro{PSD}{Positive Semidefinite}


\acro{SDP}{Semidefinite Program}
\acroindefinite{SDP}{an}{a}

\acro{SOS}{Sum of Squares}
\acroindefinite{SOS}{an}{a}


\end{acronym}

\subsection{Notation}

The set of $n$-dimensional multi-indices with sum $\leq d$ is $\N^n_d$. The symbol $\odot$ denotes the Hadamard product between two vectors. For any polynomial $h \in \R[x]$, the degree of $h$ is $\deg h$. 
Monomials are represented in multi-index form as $\forall \alpha \in \N^n, x \in \R^n: x^\alpha = \prod_{i=1}^n x_i^{\alpha_i}$. For a symmetric matrix $Q$, the symbol $Q \succeq 0$ denotes that $Q$ is \ac{PSD}. The set of nonnegative measures supported over a set $X$ is $\Mp{X}$. 
For any function $h$ continuous over $X$ and measure $\mu \in \Mp{X}$, the pairing $\inp{h}{\mu}$  represents Lebesgue integration $\int_X h(x) d \mu(x)$. This pairing forms an inner product between nonnegative continuous functions and nonnegative measures when $X$ is compact. The mass of $\mu$ is $\mu(X) = \inp{1}{\mu}$, and $\mu$ is a probability measure if this mass is 1. One distinguished probability measure is the Dirac delta $\delta_{x=x'}$ supported at the point $x' \in X$, in which $\forall f$ continuous over $X: \inp{h(x)}{\delta_{x'}} = h(x')$.



\subsection{Sum-of-Ratios Optimization}

A sum-of-ratios optimization problem with $N$ terms in a variable $x \in X \subseteq \R^n$ has the expression of \citep{schaible2003fractional}
\begin{equation}
\label{eq:sum_of_rational}
    P^* = \inf_{x \in X} \sum_{\ell=1}^L \frac{p_\ell(x)}{q_\ell(x)},
\end{equation}
for a finite number of bounded-degree polynomials $\{p_\ell, q_\ell\}$. A sufficient condition for the optimal value $P^*$ of \eqref{eq:sum_of_rational} to be finite is if $X$ is compact and $\forall x \in X$,  $\forall \ell \in \{1, \dots, L\}: q_\ell(x) > 0$. Problem \eqref{eq:sum_of_rational} is NP-complete in general \citep{freund2001solving}. 
It can be lifted into an infinite-dimensional \ac{LP} in a single probability measure $\mu \in \Mp{X}$ to form
\begin{align}
\label{eq:sum_of_rational_meas}
    p^* &= \inf_{\mu \in \Mp{X}} \sum_{\ell=1}^L \left \langle \frac{p_\ell(x)}{q_\ell(x)}\;,\; {\mu(x)} \right \rangle:& \inp{1}{\mu}=1. 
\end{align}

\cite{bugarin2016minimizing} introduces measures $\nu_\ell \in \Mp{X}$  under the relation
\begin{align}
    \forall \alpha \in \N^n: \inp{x^\alpha}{\nu_\ell(x) q_\ell(x)} = \inp{x^\alpha}{\mu(x)}. \label{eq:abscont}
\end{align}
This enforces that the measure $\nu_\ell$ is absolutely continuous with respect to $\mu$ with a density function $1/q_\ell$.
Assuming that $q_\ell(x) > 0$ over $X$, an equivalent formulation of \eqref{eq:sum_of_rational_meas} under the relations in \eqref{eq:abscont} is
\begin{subequations}
\label{eq:sum_of_rational_meas_abscont}
\begin{align}
    p^* = \inf_{\mu, \nu_\ell \in \Mp{X}} \textstyle &\sum_{\ell=1}^L \inp{p_\ell(x)}{\nu_\ell(x)}: \\
   \  & \inp{1}{\mu}=1 \\
    & \forall \ell \in\{1,\dots,L\}: \textrm{\eqref{eq:abscont} holds}.
\end{align}
\end{subequations}

Theorem 2.1 of \cite{bugarin2016minimizing} states that positivity of $q_\ell(x)$ over $X$ and finiteness of $P^*$ implies that $p^* = P^*$.

The moment-\ac{SOS} hierarchy of \acp{SDP} \citep{lasserre2009moments} may be used to convert the infinite-dimensional \ac{LP}  \eqref{eq:sum_of_rational_meas_abscont} into a set of finite-dimensional \acp{SDP}. For every $\alpha \in \N^n,$ the value $m_\alpha = \inp{x^\alpha}{\mu}$ is the $\alpha$-moment of $\mu$. 
To each measure $\mu$ there exists a unique moment-substituting linear operator $\mathbb{L}_m: \R[x] \rightarrow \R$, known as the Riesz operator, that operates $\forall p \in \R[x]$ as
\begin{align}
\label{eq:riesz_operator}
    \textstyle \mathbb{L}_m(\sum_{\alpha} p_\alpha x^\alpha) = \sum_{\alpha} p_\alpha \mathbb{L}_m(x^\alpha) = \sum_{\alpha} p_\alpha m_\alpha. 
\end{align}
Letting $m^\ell_\alpha = \inp{x^\alpha}{\nu_\ell}$ be the sequence of $\alpha$-moments for each measure $\nu_\ell$, the constraint \eqref{eq:abscont} can be interpreted as a semi-infinite set of linear equality constraints between $\{m_\alpha\}$ and each $\{m^\ell_\alpha\}$ using the respective Riesz operators. The \ac{LP} in \eqref{eq:sum_of_rational_meas_abscont} can be reformulated as an optimization problem over infinite sequences $(\{y_\alpha\}, \{y^\ell_\alpha\})$ (referred to as pseudo-moments) under the constraint that there exists a (representing) measure $\mu \in \Mp{X}$ with $\inp{x^\alpha}{\mu} = y_\alpha$.

The moment-\ac{SOS} hierarchy is applicable when the set $X$ has a particular semi-algebraic representation. \Iac{BSA} set is a set formed by the intersection of a finite number of polynomial inequalities. The general form of \iac{BSA} set in $\R^n$ with $N_c$ constraints is:
\begin{equation}
    \K = \{x \in \R^n \mid g_k(x) \geq 0, \ \forall k=1, \dots, N_c\}. \label{eq:bsa}
\end{equation}
We now define what is known as the localizing matrix. 
For a polynomial $h(x) = \sum_\gamma h_\gamma x^\gamma$  and a sequence $y = \{y_\alpha\}$, with $\gamma, \alpha \in \N^n$, the localizing matrix of order $d$ associated with $h$ and $y$ is a real symmetric matrix denoted as $\M_d[h y]$ with rows and columns indexed by $\N^n_d$, and whose entry $(\alpha, \beta)$ is given by
\begin{align}
    \textstyle \M_d[h y]_{\alpha, \beta} = \sum_{\gamma} h_\gamma y_{\alpha+\beta+\gamma}, \qquad \forall \alpha, \beta \in \N^n_d.
\end{align}


Let $\M_d[\K y]$ be the block-diagonal matrix formed by  $\M_d[1 y]$ and $\M_d[g_k y]$ for all $k\in \{1, \dots, N_c\}$.
The degree-$d$ truncation of \eqref{eq:sum_of_rational_meas_abscont} with moment sequences $y$ for $\mu$ and $y^\ell$ for $\nu^\ell$ (assuming that $X$ is \iac{BSA} set) is:
\begin{subequations}
\label{eq:sum_of_rational_mom_abscont}
\begin{align}
    p^*_d =& \inf_{y, y^\ell} \textstyle  \sum_{\ell=1}^L \mathbb{L}_{y^\ell}(p_\ell(x)): \\
     & y_\0 = 1, \\
    & \forall \ell \in \{1, \dots, L\}, \alpha \in \N^n_{2d} \nonumber \\
    & \qquad \mathbb{L}_{y^\ell}(x^\alpha q_\ell(x)) = \mathbb{L}_{y}(x^\alpha) \\
    &  \M_d[X y], \forall \ell: \M_{d + \lceil \deg q_\ell/2 \rceil }[X y^\ell] \succeq 0.\label{eq:sum_of_rational_mom_abscont_psd}
\end{align}
\end{subequations}

Constraint \eqref{eq:sum_of_rational_mom_abscont_psd} is a sufficient condition for $y$ and $\{y_\ell\}$ to have representing measures supported in $X$. This sufficient criterion is also necessary if $X$ satisfies a {\textit{compactness representation condition}:} there exists an $R>0$ such that $R - \norm{x}_2^2$ can be added to the constraint-defining functions $\{g_k(x)\}$ without changing $X$. Convergence of \eqref{eq:sum_of_rational_mom_abscont} can be established by the following theorem.
\begin{thm}[Theorem 2.2 of \cite{bugarin2016minimizing}] 
\label{thm:bugarin_lmi}
    If  $X$ satisfies the compactness representation condition, then $\lim_{d \rightarrow \infty} p^*_d = p^*$. 
Furthermore, if there exists $d, r \in \N$ such that the following conditions (known as flat extension \citep{curto1996solution}) holds for the solution of \eqref{eq:sum_of_rational_meas_abscont}:
\begin{subequations}
\label{eq:flat_extension}
\begin{align}
    r &= \rank{\M_d[1y]}\\
    r &= \rank{\M_{d - \max_k \lceil\deg{g_k}/2 \rceil}[1y]} \\
    \forall \ell: \quad r&=  \rank{\M_{d+\lceil \deg q_\ell/2\rceil}[1y^\ell]} \\
    \forall \ell: \quad r &= \rank{\M_{{d+\lceil \deg q_\ell/2\rceil} - \max_k \lceil\deg{g_k}/2 \rceil}[1y^\ell]},
\end{align}
\end{subequations}
then there exists a representing measure $\mu \in \Mp{X}$ for $y$ supported at $r$ distinct points in $X$, and each one of these support points is globally optimal for \eqref{eq:sum_of_rational}.


\end{thm}

\smallskip
\begin{rem}
    The formulation in \cite{bugarin2016minimizing} only involves $L$ measures $\nu^\ell$. The presentation given in this introduction added a new measure $\mu$ in order to aid in the explanation.

\end{rem}

\section{Sum-of-Rational System Identification}
\label{sec:rational_sysid}

We now pose \eqref{eq:gain_base} as a sum-of-rational optimization problem, and show how it can be lifted into an infinite-dimensional  \ac{LP} over $\Mp{X}$. We will require the following assumptions:

\begin{assum}\label{assum1}
    There exists \iac{BSA} set $\K_0$ satisfying the compactness representation condition such that $(a^*, b^*) \in \K_0$
\end{assum}
\vspace{1pt}
\begin{assum}\label{assum2}
    There exists $(a', b') \in \K_0$ such that $\forall f: 1+
    \sum_{k=1}^n a'_k e^{-k j \omega_f}
   \neq 0$.
\end{assum}



\subsection{Support Sets}

The set $\K_0$ in Assumption \ref{assum1} contains a priori known boundedness information on the magnitude/values of the parameter vectors $(a, b)$ that could have generated the observed data in $\dc$. 

Define $\mathcal{S}$ as the set of Schur-stable plants $(a, b)$, through the use of a Hermite-Matrix \ac{PMI} \citep{1593895}.
Following the procedure of \cite{henrion2011inner}, define the symmetric-matrix-valued polynomial
\begin{align}
    \Xi(a) := \Theta(a)^T \Theta(a) - \tilde{\Theta}(a)^T \tilde{\Theta}(a),
\end{align}
where $\Theta(a)\in\R^{n\times n}$ and $\tilde{\Theta}(a)\in\R^{n\times n}$ are defined as
\begin{align}
 \!\!\!\!   \Theta(a) \!=\! \begin{bmatrix}
        1 & a_1 & a_2 & \hdots\\
        0 & 1 & a_1 & \hdots \\
        0 & 0 & 1 & \hdots \\
        \vdots & \vdots & \vdots & \ddots
    \end{bmatrix}\!\!,\   \tilde{\Theta}(a)\! =\! \begin{bmatrix}
        a_n & a_{n-1} & a_{n-2} & \hdots\\
        0 & a_n & a_{n-1} & \hdots \\
        0 & 0 & a_n & \hdots \\
        \vdots & \vdots & \vdots & \ddots
    \end{bmatrix}\!\!,
    \end{align}
respectively where $n$ here is the model order.
The set of plants $(a, b)$ such that $z^n (1+A(z^{-1}))$ has roots strictly inside the unit disk can be expressed as 
\begin{align}
    \mathcal{S} = \{(a, b) \in \R^{2n}: \qquad  \Xi(a) \succ 0\}. \label{eq:stable_pmi_orig}
\end{align}

The open set \eqref{eq:stable_pmi_orig} has a positive-definite matrix description. For a given small tolerance $\delta > 0$, we form the closed set $\mathcal{S}_\delta \subset \mathcal{S}$ defined by 
\begin{align}
    \mathcal{S}_\delta = \{(a, b) \in \R^{2n}: \qquad  \Xi(a) \succeq \delta I_{n}\}. \label{eq:stable_pmi}
\end{align}
The set of Schur-stable plants with a stability margin of $\delta$ that are consistent with the known parameter bounds is 
\begin{align}
    \K_s = \K_0 \cap \mathcal{S}_\delta.
\end{align}

\begin{rem}
\label{rmk:compact_delta}
    The $\delta$-stability margin ensures that $\K_s$ is compact (closed and bounded by Heine-Borel) under Assumption \ref{assum1}.
\end{rem}

\begin{rem}
\ac{BSA} sets  \eqref{eq:bsa} can also include \ac{PMI} constraints $Q(x)  \succeq 0$. See \cite{1593895} for its explicit representation and the associated localizing matrices.
\end{rem}


The symbol $\K$ will refer to $\K_0$ or $\K_s$ as appropriate. We will specifically highlight the differences between properties of sum-of-rational identification problems that are posed over $\K_0$ versus $\K_s$. In the following, we will assume that $\K_0$ could potentially include Schur-unstable plants.




\subsection{Sum-of-Rational Formulation}

We can express the per-frequency residual $\varepsilon_f$ as the following rational expression:
\begin{subequations}
\label{eq:per_residual}
\begin{align}
    \varepsilon_f &= G_f - G(e^{j \omega_f}) \\
    &= G_f - \frac{B(e^{-j \omega_f})}{1 + A(e^{-j \omega_f})} \\
    &= \frac{G_f(1 + A(e^{-j \omega_f})) - B(e^{-j \omega_f})}{1 + A(e^{-j \omega_f})}.
\end{align}
\end{subequations}

The per-frequency rational error term $\frac{p_f(a, b)}{q_f(a)}$ for \eqref{eq:gain_base} can be described from \eqref{eq:per_residual} by
\begin{subequations}
\label{eq:poly_base}
\begin{align}
    p_f(a, b) &= \abs{W_f G_f(1 + A(e^{-j \omega_f})) - W_f B(e^{-j \omega_f})}^2 \label{eq:poly_base_num}\\
    q_f(a) &= \abs{1 + A(e^{-j \omega_f})}^2. \label{eq:poly_base_q}
\end{align}
\end{subequations}

The polynomials in \eqref{eq:poly_base} satisfy $p_f \in \R[a, b]$, $q_f \in \R[a]$ and are therefore real-valued (given that $(a, b)$ are real parameters). Both $p_f$ and $q_f$ are quadratic in $(a, b)$ for each frequency point $f \in \{1, \ldots, N_f\}.$

Problem \eqref{eq:gain_base} can therefore be reformulated as
\begin{align}
    P^* = \inf_{(a, b)} \mathcal{J}_\dc(a,b) = \inf_{(a, b) \in \K} \sum_{f=1}^{N_f} \frac{p_f(a, b)}{q_f(a)}. \label{eq:gain_rat}
\end{align}
\smallskip

\begin{lem}
    Each function $q_f(a)$ is nonnegative over $\K_0$.
    \label{lem:nonneg_unstab}
\end{lem}
\begin{pf}
    The function $q_f(a)$ from \eqref{eq:poly_base} arises from a squared-norm expression, and is therefore nonnegative for every $(a, b) \in \R^{2n} \supset \K_0$. 
\end{pf}
\smallskip

\begin{cor}
The objective of \eqref{eq:gain_rat} satisfies $P^* \geq 0$, given that \eqref{eq:gain_base} infimizes the sum of squared-norms.
\end{cor}
\smallskip

\begin{lem}
    Each function $q_f(a)$ is positive over $\K_s$.
    \label{lem:q_pos_stab}
\end{lem}
\begin{pf}
    The function $q_f(a)$ will be zero if $e^{-j \omega_f}$ is a root of $1 + A(z^{-1})$. Given that $(a, b) \in \mathcal{S}$ requires that all roots of $1 + A(z^{-1})$ have magnitudes less than one, $e^{-j\omega_f}$ cannot be a root of $1 + A(z^{-1})$. As a result, $q_f(a)$ is positive over $\K_s$ for each $f$. 
\end{pf}


\subsection{Measure Linear Program}


The sum-of-rational program \eqref{eq:gain_rat} can be converted into a measure \ac{LP} through the introduction of a probability distribution $\mu \in \Mp{\K}$ and a sequence of per-frequency measures $\forall f \in \{1, \ldots, N_f\}: \ \nu_f \in \Mp{\K}$.

Application of the sum-of-rational framework from \citep{bugarin2016minimizing} in \eqref{eq:sum_of_rational_meas_abscont} towards \eqref{eq:gain_rat} results in
\begin{subequations}
\label{eq:rational_meas_sparse}
\begin{align}
    p^* = \textstyle  \inf & \textstyle \sum_{f=1}^{N_f} \inp{p_f(a, b)}{\nu_f(a, b)} \label{eq:rational_meas_sparse_obj} \\
    \textrm{s.t.} \quad & \inp{1}{\mu} = 1 \label{eq:rational_meas_sparse_prob} \\
    & \forall (\alpha, \beta) \in \N^{2n}, f \in \{1, \ldots, N_f\}: \nonumber \\
    & \qquad \inp{a^\alpha b^\beta q_f(a)}{\nu_f (a, b)} = \inp{a^\alpha b^\beta}{\mu(a, b)} \label{eq:rational_meas_sparse_mult}\\
    & \mu \in \Mp{\K} \label{eq:rational_meas_sparse_mu}\\
    & \forall f \in \{1, \ldots, N_f\}: \ \nu_f \in \Mp{\K}.
\end{align}
\end{subequations}
\smallskip

\begin{thm}
\label{thm:meas_lower_bound}
    Program \eqref{eq:rational_meas_sparse} lower-bounds \eqref{eq:gain_base} with $p^* \leq P^*$ over $\K_0$ under Assumptions \ref{assum1}-\ref{assum2}.
\end{thm}
\begin{pf}
    Let $(a', b') \in \K$ be a feasible model (exists by Assumption \ref{assum2}). Then we can choose $\mu' = \delta_{a=a', b=b'}$, and also pick $\nu_f' = (\1/q_f(a'))\delta_{a=a', b=b'}$. Assumption \ref{assum2} ensures that $q_f(a') > 0$, which produces a bounded $\nu_f'$. The objective with $\nu_f'$ is \begin{align}
   \sum_{f=1}^{N_f} \Big \langle {p_f(a, b)}\;, \;{\frac{1}{q_f(a')}\delta_{a=a', b=b'}} \Big \rangle &= \sum_{f=1}^{N_f}  \frac{p_f(a', b')}{q_f(a')} \nonumber  \\
        &= \mathcal{J}_\dc(a', b').
    \end{align}


We now consider the case where a plant can have a pole at one of the sampled frequencies $e^{-j \omega}$.
Let $\mathcal{F} \subseteq \{1, \ldots, N_f\}$ with $\abs{\mathcal{F}} \geq 0$ be a set of sampling indices. 
Assume there exists a plant $(\bar{a}, \bar{b})$ with polynomials $\bar{A}(z) =  \sum_{i=1}^n \bar{a}_i z^i$ and $\bar{B}(z) = \sum_{i=1}^n \bar{b}_i z^i$ such that $\forall f \in \mathcal{F}: 1+\bar{A}(e^{-j \omega_f}) = 0$. It therefore holds that $q_f(\bar{a})=0$ for every $f \in \mathcal{F}$. The measures $\nu_f$ with $f \in \mathcal{F}$ are then unrestricted by \eqref{eq:rational_meas_sparse_mult}. Given that $p_f(a, b) \geq 0$ over $\R^{2n}$ (via the norm constraint), it holds that the minimum possible value for $\inp{p_f}{\nu_f}$ is $0$ by choosing $\nu_f =0$. All other measures $\nu_{f'}$ for $f' \not\in \mathcal{F}$ can be chosen as $\nu_{f'}' = (1/q_{f'}(\bar{a})) \delta_{a=\bar{a}, b=\bar{b}}.$ The objective in \eqref{eq:rational_meas_sparse_obj} only contributes  points $f' \not\in \mathcal{F}$.

The numerator satisfies $p_f(\bar{a}, \bar{b}) = \bar{B}(e^{-j\omega_f})$ when $(1+\bar{A}(e^{-j\omega f})) = 0$. For every $f \in \mathcal{F}$, let $m_f$ denote the multiplicity of the root $e^{j \omega_f}$ in $(1+\bar{A}(z))$.
There are two possible circumstances going forward. The first is that $\bar{B}(z)$ also has a root at $e^{-j \omega_f}$ with multiplicity $m_f' \geq m_f$ (removable singularity), which implies the existence of a complete pole-zero cancellation leaving $p_f(\bar{a}, \bar{b})/q_f(\bar{a})$ finite (with $p_f(\bar{a}, \bar{b})/q_f(\bar{a})=0$ if $m_f' > m_f$). The second circumstance is that $m_f'<m_f$ (including if $\bar{B}(e^{-j \omega_f}) \neq 0$), which preserves the pole at $e^{-j \omega_f}$ leading to $p_f(\bar{a}, \bar{b})/q_f(\bar{a}) = \infty$.


Considering that $\forall f \in \mathcal{F}: \ \nu_f=0$, the objective in \eqref{eq:rational_meas_sparse_obj} would return 
\begin{align}
        \textstyle \sum_{f' \notin \mathcal{F}}^{N_f} \inp{p_{f'}(a, b)}{\nu'_{f'}} &= \textstyle \sum_{f' \notin \mathcal{F}}^{N_f} p_f(a', b')/q_f(a') \nonumber  \\
        &\leq  \mathcal{J}_\dc(a', b').
    \end{align}

In summary, the measures $\mu$ and $\nu_f$ may be constructed for every $(a, b) \in \K_0$ with an objective in \eqref{eq:rational_meas_sparse_obj} attaining the $\mathcal{J}_\dc(a', b')$ ($\abs{\mathcal{F}} = 0$) or lower-bounding it ($\abs{\mathcal{F}} > 0$). This proves the theorem.
\end{pf}
\smallskip

\begin{thm}
    Program \eqref{eq:rational_meas_sparse} has the same optimal value as \eqref{eq:gain_rat} with $p^* = P^*$ under A1-A2 and over $\K_s$.
\end{thm}
\begin{pf}
    The bound $p^* \leq P^*$ is proven by Theorem \ref{thm:meas_lower_bound} in the circumstance where $e^{-j \omega_f}$ is not a pole of $1+A(z^{-1})$. Because each $q_f(a)$ is positive over $\K_s$ (Lemma \ref{lem:q_pos_stab}) and $\K_s$ satisfies the compactness condition (A1), Theorem 2.1 of  \citep{bugarin2016minimizing} certifies that $p^* = P^*$.
\end{pf}

\section{Sum-of-Rational Semidefinite Program}

\label{sec:mom_sdp}

This section analyzes finite-degree moment-\ac{SOS} truncations of \eqref{eq:rational_meas_sparse}. The main results are:
\begin{itemize}
    \item Problems posed 
over $\K_s$ have a guarantee of convergence in objective to the minimal \ac{MSE}. 
\item Problems posed over $\K_0$ lack this guarantee of convergence, but require fewer computational resources to solve as compared to those over $\K_s$.
\end{itemize}

\subsection{LMI and Convergence}
A degree-$d$ truncation of \eqref{eq:rational_meas_sparse} with pseudomoment sequences $(y, y^f)$ for $(\mu, \nu_f)$ from \eqref{eq:sum_of_rational_mom_abscont} is
\begin{subequations}
\label{eq:sum_of_rational_lmi}
\begin{align}
\label{eq:sum_of_rational_mom_abscont_sysid}
    p^*_d =& \inf_{y, y^f} \textstyle  \sum_{f=1}^L \mathbb{L}_{y^f}(p_f(a, b)) \\
 \textrm{s.t.} \quad    & y_\0 = 1 \\
    & \forall f \in \{1, \ldots, N_f\}, (\alpha, \beta) \in \N^n_{2d} \nonumber \\
    &\qquad \mathbb{L}_{y^f}(a^\alpha b^\beta q_f(a))  =   \mathbb{L}_{y}(a^\alpha b^\beta) \label{eq:sum_of_rational_mom_abscont_marg} \\
    & \M_d[\K y], \forall f: \M_{d + 1 }[\K y^f] \succeq 0\label{eq:sum_of_rational_mom_abscont_mom}.
\end{align}
\end{subequations}


\smallskip

\begin{cor}
    The sequence of objectives in \eqref{eq:sum_of_rational_lmi} satisfy $p^*_k \leq p^*_{d+1}$ for all $d$ under Assumptions \ref{assum1}-\ref{assum2} over $\K_0$.
\end{cor}
\begin{pf}
The sequence $\{p^*_d\}$ is monotonically nondecreasing, because the number of truncated moments considered in the infinite-dimensional matrix constraints \citep{lasserre2009moments} is increased at each $d$.
\end{pf}

\smallskip
\begin{thm}
    The sequence of objectives in \eqref{eq:sum_of_rational_lmi} converge to $\lim_{d \rightarrow \infty} p^*_d = P^*$ under Assumptions \ref{assum1}-\ref{assum2} over $\K_s$.
\end{thm}
\begin{pf}
    All conditions of Theorem \ref{thm:bugarin_lmi} (Theorem 2.2 of \cite{bugarin2016minimizing}) are met (compactness of $\K_s$ and positivity of $q_f$ over $\K_s$), thus affirming convergence.
\end{pf}

\subsection{Computational Complexity}

The computational complexity of solving a moment-\ac{SOS} \ac{LMI} of the form in \eqref{eq:sum_of_rational_mom_abscont} with $N$ variables, a degree $d$, single moment matrices $\M_d[1y]$ of size $\binom{N+d}{d}$, and $L = 0 $ using an interior-point \ac{SDP} (up to $\epsilon$-accuracy) is $O(N^{6d})$ or $O(k^{dN})$ \citep{lasserre2009moments}.

Problem \eqref{eq:gain_rat} is an optimization problem posed over the $2n$ variables $(a, b)$. Table \ref{tab:mom_size} records the sizes and multiplicities of the matrix variables in \eqref{eq:sum_of_rational_lmi} when the full monomial basis is used to describe constraints.

\begin{table}[h]
    \centering
    \caption{Size and multiplicity of \ac{PSD} matrix variables in \ac{LMI} \eqref{eq:sum_of_rational_mom_abscont_sysid}}
    \begin{tabular}{c l l l c}
    Measure & Matrix &  Size ($\K_0$) & Size ($\K_s$) & Mult. \\ \hline
        $\mu$ & $\M_d[\K y]$ & $\binom{2n+d}{d}$ & $n\binom{2n+d-1}{d-1}$& 1 \\
        $\nu_f$ & $\M_{d+1}[\K y^f]$ & $\binom{2n + d+1}{d+1}$ & $n\binom{2n+d}{k}$& $N_f$
    \end{tabular}
    \label{tab:mom_size}
\end{table}

The largest \ac{PSD} matrix for each measure over $\K_0$ is the moment matrix $(\M_d[y], \M_{d+1}[y^f])$. The largest \ac{PSD} block for $\K_s$ occurs in the stabilizing-\ac{PMI} constraint \eqref{eq:stable_pmi} with localizing matrix $(\M_d[\mathcal{S} y], \M_{d+1}[\mathcal{S} y^f])$. 

Notice that the sizes of the matrices do not depend on $N_f$. The computational scaling for fixed $n$ therefore grows exponentially in $d$ and linearly in $N_f$, according to the dominant \ac{PSD} blocks for $\nu_f$ of size $\binom{2n+d+1}{2n+1}$ ($\K_0$) or $n\binom{2n+d}{d}$ ($\K_s$). By construction, the non-\ac{PMI} localizing matrix blocks (such as $\M_{d+1}[g_j(a, b) y^f] \succeq 0$ for a describing constraint polynomial $g_j(a, b)$ of $\K$) will have a smaller size than the moment block $\M_{d+1}[1y^f]$.


\subsection{Recovery of Optimal Plant}

\label{sec:recovery}


Suppose that \eqref{eq:sum_of_rational_mom_abscont_sysid} is solved at degree $d$ to produce an objective $p^*_d \leq P^*$ and moment matrix $\M_d[1y]$. Further suppose that the moment matrix $\M_d[1y]$ satisfies \eqref{eq:flat_extension} with rank $r'$ (up to some numerical accuracy) and has support points $\{(a^*_c, b^*_c)\}_{c=1}^{r'}$. Then, the support points can be extracted via the numerical algorithm proposed in \cite{henrion2005detecting}. If there exists a sufficiently small $\epsilon \geq 0$ such that  one of the support points $c$ satisfies $\mathcal{J}_\dc(a^*_c, b^*_c) \in [p^*_d, p^*_d + \epsilon]$, then the point $(a^*_c, b^*_c)$ is approximately optimal. 

\section{Numerical Examples}

\label{sec:examples}

Code to perform all the simulation experiments in this section is available online\footnote{\href{https://doi.org/10.3929/ethz-b-000641430}{https://doi.org/10.3929/ethz-b-000641430}}. The sum-of-rational \acp{SDP} (without stability constraints) were formulated using Gloptipoly \citep{henrion2003gloptipoly}, modeled through YALMIP \citep{lofberg2004yalmip}, and solved through Mosek 10.1 \citep{mosek92}. Stability-constrained sum-of-rational system identification was written in Julia through a Correlative-Term-Sparsity interface (CS-TSSOS) \citep{wang2022cs}, and converted into Mosek using \texttt{JuMP} \citep{Lubin2023jump}.

Each example involves identification of a discrete-time system whose complex gain values are measured at the frequency points  $\omega_f = \frac{(f-1)\pi}{10}$ for all $f \in \{1, \ldots, 11\}$.
The data $G_f$ from \eqref{eq:gain_datas} is corrupted by additive noise $\eta_f$ drawn from a circular complex normal distribution with zero-mean and standard deviation 0.3 when $f \in \{2, \ldots 10\}$. The data points $G_1$ and $G_{11}$ (frequencies $0$ and $\pi$) are corrupted by real zero-mean normally distributed noise with standard deviation 0.3. This type of noise model corresponds to an empirical transfer function estimate obtained via periodic time-domain data when the time-domain noise is i.i.d. normal distributed white noise.
Weights of $W_f = 1$ are used for each frequency point $\omega_f$.

\subsection{Fixed Third-Order System}

This example involves the identification of a third-order  model with a ground-truth system
\begin{equation}
        G_\circ(z) = \frac{2 z^{-1} - z^{-3}}{1 - 0.18 z^{-1} - 0.134 z^{-2} - 0.637 z^{-3}}. \label{eq:truth_3}  
\end{equation}

System \eqref{eq:truth_3} has stable poles at $z = 0.98$ and $z = -0.4 \pm j0.7$. \textit{Stability is enforced}, and $\K_s$ is defined with the box constraint $\norm{[a^\top\ b^\top]}_\infty \leq 2$ and $\delta = 10^{-4}$.  Table \ref{tab:ex1_bounds} lists objective values obtained by applying the proposed method \eqref{eq:sum_of_rational_lmi}, as well as by the \textsc{Matlab} functions \texttt{n4sid}, \texttt{oe} and \texttt{fmincon} initialized at the \texttt{n4sid} solution. 

\begin{table}[h]
    \centering
    \caption{Objective values $\mathcal{J}_\dc(a,b)$ for  \eqref{eq:truth_3}}
\begin{tabular}{c|ccccc}
     Method& \eqref{eq:sum_of_rational_lmi} at $k=1$ & \texttt{n4sid}&  \texttt{oe}& \texttt{fmincon(n4sid)} \\
     $\mathcal{J}_\dc(a,b)$ & 0.3173 & 0.3368 & 0.3224 & 0.3173
\end{tabular}    
    \label{tab:ex1_bounds}
\end{table}

The identified system extracted from the solved moment matrix $\M_1[1y]$ by the method of Section \ref{sec:recovery} is:
\begin{equation}
        \hat{G}(z) = \frac{1.978 z^{-1} +0.0835 z^{-2} - 1.1090 z^{-3}}{1 - 0.1415 z^{-1} - 0.2016 z^{-2} - 0.6107 z^{-3}}. \label{eq:mse_3}  
\end{equation}

The recovered system  \eqref{eq:mse_3} has an evaluated objective value of 0.3173 (upper-bound). Given that our method \eqref{eq:sum_of_rational_lmi} at $k=1$ (in Table \ref{tab:ex1_bounds}) returns a lower-bound MSE of 0.3173 over $\K_s$, it holds that system \eqref{eq:mse_3} is \textit{globally optimal} over $\K_s$.
System \eqref{eq:mse_3} is also recovered by \texttt{fmincon(n4sid)} as a local optimum without verification of global optimality or enforcing stability; our method certifies that this local optimum is a global optimum within $\K_s$. Using \texttt{fmincon} with an initial point of $a=\0, b=\0$ did not converge to a solution after 3000 function evaluations, and produced an MSE of 31.540 at iteration 3000. 

\begin{figure}[ht]
    \centering
    \includegraphics[width=0.9\linewidth]{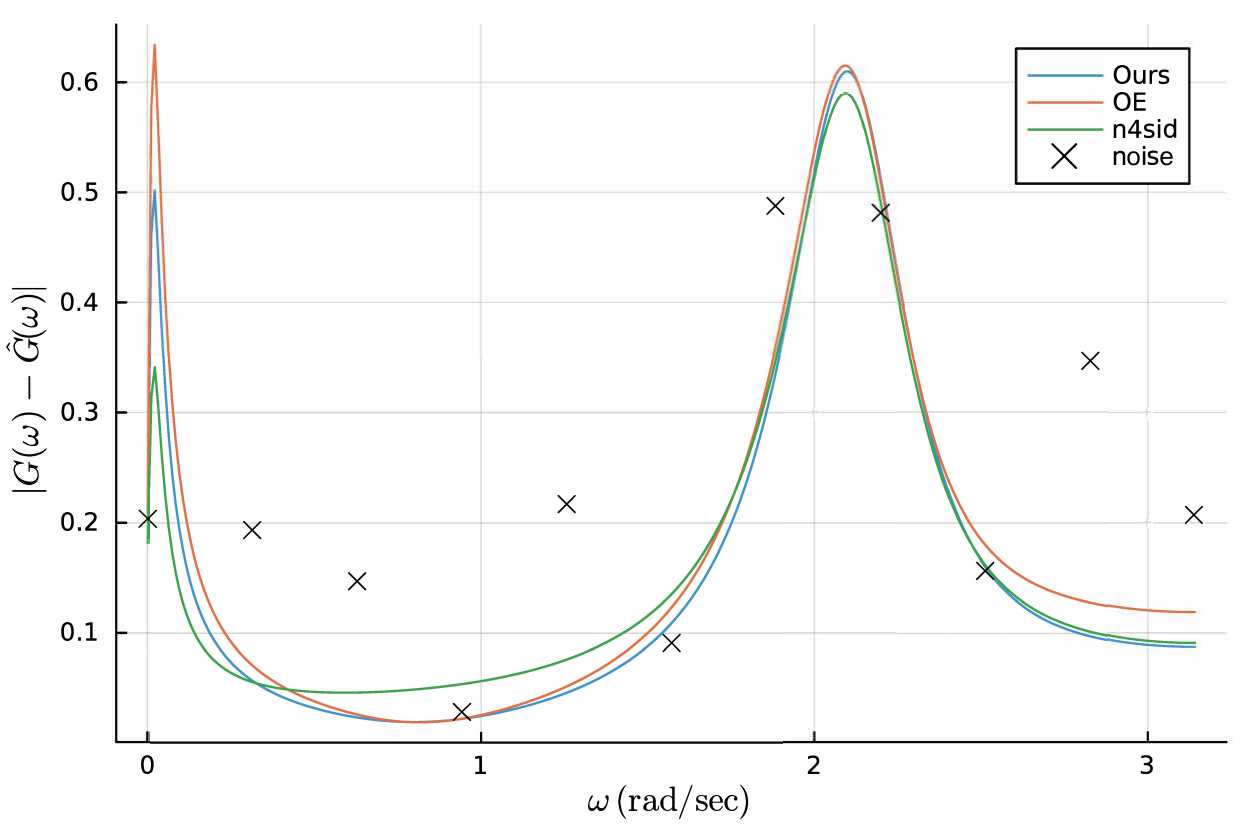}
    \caption{\label{fig:error_3} Gain errors for the different methods}
\end{figure}

Figure \ref{fig:error_3} plots the error between the frequency responses of the identified systems and the true systems.





\subsection{Monte-Carlo Second-Order Systems}
\begin{center}
\begin{figure}[ht]
\centering
\includegraphics[width=0.9\linewidth]{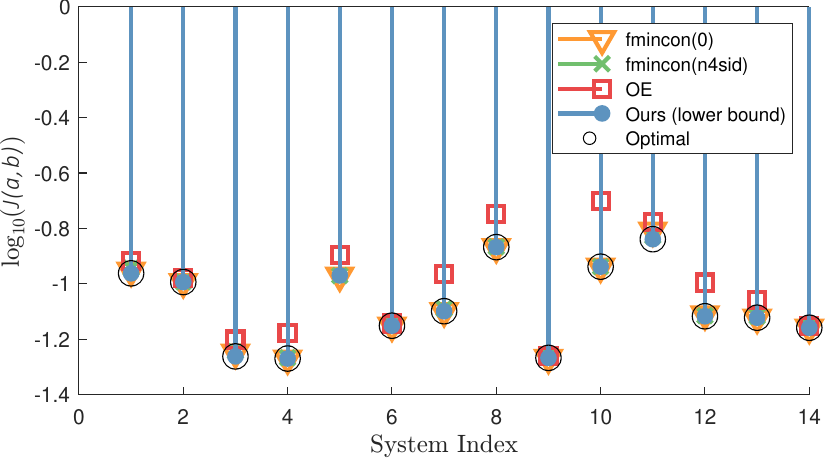}
    \caption{Comparison of objective function values in Monte-Carlo tests. The objective values obtained by \texttt{fmincon(n4sid)} are close to the  values obtained by the proposed method}
    \label{fig:rat_sysid_out}
\end{figure}
\end{center}

This example involves the identification of 14 randomly generated Schur-stable second-order systems. Figure \ref{fig:rat_sysid_out} compares the MSE of \texttt{fmincon} starting at 0, \texttt{fmincon} starting at the N4SID estimate, the \texttt{oe} method, and our approach at $k=3$ (lower-bound SDP value from \eqref{eq:sum_of_rational_lmi}). The black circles correspond to systems in which a near-optimal system is extracted by the procedure in Section \ref{sec:recovery}. A box constraint of $\norm{[a^\top\ b^\top]}_\infty \leq 1.6$ is used on parameters, and stability is not enforced.

\section{Conclusion}

\label{sec:conclusion}

We proposed a method of estimating canonical rational transfer functions based on the global minimization of a frequency-response weighted residuals 2-norm objective. The original non-convex problem is converted to an equivalent linear program in measures by using a sum-of-rational formulation. Certifiable global optima can be obtained through moment-\ac{SOS}-based truncations of the linear program in measures.



The extended version of this paper at \cite{abdalmoaty2023rationalsysid} includes further discussions about strong duality, identification of continuous-time systems,  closed-loop identification, and identification of outlier-corrupted data.
Future work includes 
the addition of regularization terms in order to prevent overfitting the noise process, and a homogenized formulation to reduce the conservatism of Assumption \ref{assum1}. Other work includes improvement of the \ac{SDP}'s numerical conditioning, especially for cases where the poles are close to the boundary of the unit disk.


\bibliography{references}

                                    
\end{document}